\documentclass[11pt,a4paper,american]{extarticle}

\usepackage[T1]{fontenc}
\usepackage[latin1]{inputenc}
\usepackage{babel}
\usepackage{amsmath}
\usepackage{amssymb}
\usepackage{setspace}
\usepackage{esint}
\usepackage[numbers]{natbib}
\usepackage{xcolor}
\usepackage{babel}
\usepackage{bbm}
\usepackage{hyperref}
\usepackage{authblk}
\usepackage{tikz}
\usetikzlibrary{matrix}
\usepackage{cancel}
\usepackage{comment}

\allowdisplaybreaks
\makeatletter
\numberwithin{equation}{section}\date{}

\title{
\vspace{-4em} 
\begin{flushright}
{\small QMUL-PH-25-26}
\end{flushright}
\vspace{2em}
{\bf Tension between string amplitude prescriptions with presumed spacetime supersymmetry}
\vspace{2em}
}

\author[1,3]{Thales Azevedo\thanks{thales@if.ufrj.br}}
\affil{Instituto de F\'isica, Universidade Federal do Rio de Janeiro, \authorcr Av. Athos da Silveira Ramos 149, 21941-972, Rio de Janeiro - Brazil}

\author{Alessandro  Georgoudis\thanks{a.georgoudis@qmul.ac.uk}}
\affil{Centre for Theoretical Physics, Department of Physics and Astronomy,
Queen Mary University of London, Mile End Road, London E1 4NS, United Kingdom}

\author{Renann Lipinski Jusinskas\thanks{renannlj@fzu.cz}}
\author[3,4]{Sitender Pratap Kashyap\thanks{kashyap@fzu.cz}}
\affil{Institute of Physics of the Czech Academy of Sciences \& CEICO,
\authorcr  Na Slovance 2, 182 21, Prague - Czech Republic}

\affil[4]{ICTP - South American Institute for Fundamental Research, IFT-UNESP\\ 
\authorcr Rua Dr. Bento Teobaldo Ferraz 271, 01140-070, S\~ao Paulo, SP, Brazil}
\makeatother

\begin{document}
\maketitle
\begin{abstract}
We analyze an alternative prescription for computing tree level scattering amplitudes in the pure spinor superstring. Proposed by Berkovits, and inspired by an earlier work by Lee and Siegel, it bypasses the traditional ghost number three measure of the pure spinor ghosts, $\langle\lambda^3 \theta^5\rangle=1$, arguably simplifying the calculation. However, direct computations using the free field OPEs of the target superspace CFT fail to reproduce the expected $\textrm{SL}(2,\mathbb{C})$ invariance of tree level  correlators. This persistent mismatch, confirmed through independent methods, is not compatible with the gauge invariance of the amplitude. We also discuss possible implications for the B-RNS-GSS formalism of the superstring.
\thispagestyle{empty}
\newpage
\tableofcontents{}
\end{abstract}

\section{Introduction\label{sec:motivation}}

Early on in the development of string theory, it was clear that a minimally realistic connection to particle physics would involve supersymmetry. In the spinning string, i.e. the Ramond--Neveu--Schwarz (RNS) formalism \cite{Ramond:1971gb,Neveu:1971rx}, spacetime supersymmetry is not manifest, with a description of bosons and fermions that is inherently asymmetric. This limitation motivated the Green--Schwarz (GS) formalism \cite{Green:1980zg,Green:1981yb}, in which spacetime supersymmetry was manifest at the level of the world-sheet action. However, the quantization of the GS superstring is only possible in the light-cone gauge, i.e. the Lorentz symmetry is not manifest, and the formalism does not provide an efficient framework to handle string scattering. These obstacles were finally overcome with Berkovits' pure spinor (PS) superstring \cite{Berkovits:2000fe}. Despite lacking a formal derivation from first principles, the PS formalism has become an essential tool for state-of-the-art computations of string scattering amplitudes at tree and loop level (e.g. \cite{Mafra:2018nla,Mafra:2018pll,Mafra:2018qqe,Berkovits:2005ng,Gomez:2013sla}).

Key features of the pure spinor superstring are a bosonic string-like simplicity for computing world-sheet correlators and the manifest character of the super Poincar\'e symmetry. The latter follows from the construction of vertex operators in terms of superfields, so that bosonic and fermionic degrees of freedom are described on equal footing with full covariance. Arguably, one of the drawbacks of the PS formalism is the fact that superfields describing massive higher spin multiplets, such as the ones populating the string spectrum, are much less known. Indeed, the explicit construction of related vertex operators has not gone beyond the first massive level in the open string \cite{Berkovits:2002qx,Chakrabarti:2017vld,Chakrabarti:2018mqd}.

String amplitudes with massive, higher-spin external states have attracted renewed interest, mainly due to their relevance for modeling gravitational-wave scattering by black holes (see e.g.~\cite{Cangemi:2022abk,Azevedo:2024rrf,Firrotta:2024fvi,Alessio:2025nzd}, also \cite{Schlotterer:2010kk} for earlier results). 
Within the  PS formalism, however, existing studies of scattering amplitudes typically include at most one massive vertex operator~\cite{Chakrabarti:2018bah,Kashyap:2024qor,Mafra:2024fiy,Mafra:2025pmz}. 

Tree-level computations in the PS superstring are by now well established (see~\cite{Mafra:2022wml} for an excellent review and literature survey). These computations rely on a characteristic measure factor that mixes pure spinor ghosts with superspace variables. An alternative prescription was proposed in~\cite{Berkovits:2016xnb}, supported by earlier work~\cite{Lee:2006pa}, to compute manifestly super-Poincar\'e-invariant amplitudes while avoiding the pure spinor measure altogether. The key idea is to work directly with the integrands of the integrated vertex operators, which are ghost number zero objects and resemble bosonic string correlators. In principle, this prescription could offer a more systematic computation of the scattering of massive string states, such as the Compton scattering.\footnote{We have recently learned that C.~Huang, C.~R.~Mafra, and Y.~X.~Tao are pursuing similar results using the standard PS prescription~\cite{Mafra:2025abc}.}

In this work, we put to test this alternative prescription in the familiar case of the tree-level three-point correlator with massless external states. Unfortunately, we show that it fails to reproduce the tree-level amplitude involving one boson and two fermions. Since this outcome is somewhat unexpected, we provide extensive details of the computation. We conclude with comments on the results of~\cite{Lee:2006pa} and implications for the B-RNS-GSS formalism~\cite{Berkovits:2021xwh}. For completeness, we present in the appendix an elementary check of the computation.

\section{Review and conventions\label{sec:PSsuperstring}}

In this section we will briefly review some basic features of the
pure spinor formalism of the superstring, focusing on its holomorphic sector. 

The matter variables describe an $\mathcal{N}=1$ superspace in $D=10$. We denote by $X^{m}$ the target space coordinates of the string,
with $m=0,\ldots,9$. Their superpartners are denoted by
$\theta^{\alpha}$, with $\alpha=1,\ldots,16$, and their conjugates are denoted by $p_{\alpha}$.
These satisfy the following OPEs,\begin{subequations}\label{eq:OPEs}
\begin{align}
\partial X^{m}(z)\,X^{n}(y,\bar{y}) & \sim-\left(\frac{\alpha'}{2}\right)\frac{\eta^{mn}}{(z-y)},\\
p_{\alpha}(z)\,\theta^{\beta}(y) & \sim  \frac{\delta_{\alpha}^{\beta}}{(z-y)},
\end{align}
\end{subequations}where $\alpha'$ is the string length squared,
and $\eta^{mn}=\mathrm{diag}(-1,+1,\ldots,+1)$ is the Minkowski metric.

The supersymmetry algebra,
\begin{equation}
\{q_{\alpha},q_{\beta}\}=\frac{4\zeta}{\alpha'}\gamma_{\alpha\beta}^{m}\oint\partial X_{m},
\end{equation}
is realized through the supercharge
\begin{equation}
q_{\alpha}=\oint\{p_{\alpha}+\frac{2\zeta}{\alpha'}(\gamma_{m}\theta)_{\alpha}\partial X^{m}+\frac{\zeta^{2}}{3\alpha'}(\gamma_{m}\theta)_{\alpha}(\theta\gamma^{m}\partial\theta)\},
\end{equation}
where we keep an arbitrary constant parameter $\zeta$ to make it easier to
map different conventions in the literature. In addition to the Clifford
algebra, $\{\gamma^{m},\gamma^{n}\}=2\eta^{mn}$, the (symmetric) gamma matrices
$\gamma_{\alpha\beta}^{m}$, $\gamma^{m\alpha\beta}$   satisfy the identity
\begin{equation}
\eta_{mn}(\gamma_{\alpha\beta}^{m}\gamma_{\gamma\lambda}^{n}+\gamma_{\beta\gamma}^{m}\gamma_{\alpha\lambda}^{n}+\gamma_{\alpha\gamma}^{m}\gamma_{\beta\lambda}^{n})=0.
\end{equation}

Given that the theory is manifestly supersymmetric, it is useful to introduce the 
invariant combinations
\begin{align}
\Pi^{m} & =\partial X^{m}+\zeta(\theta\gamma^{m}\partial\theta),\\
d_{\alpha} & =p_{\alpha}-\frac{2\zeta}{\alpha'}(\gamma_{m}\theta)_{\alpha}\partial X^{m}-\frac{\zeta^{2}}{\alpha'}(\gamma_{m}\theta)_{\alpha}(\theta\gamma^{m}\partial\theta),
\end{align}
which yield the following OPEs,
\begin{align}
\Pi^{m}(z)\,\Pi^{n}(y) & \sim-\left(\frac{\alpha'}{2}\right)\frac{1}{(z-y)^{2}}\eta^{mn},\\
d_{\alpha}(z)\,\Pi^{m}(y) & \sim + \frac{2\zeta}{(z-y)}(\gamma^{m}\partial\theta)_{\alpha},\\
d_{\alpha}(z)\,d_{\beta}(y) & \sim-\frac{2\zeta}{(z-y)}\left(\frac{2}{\alpha'}\right)\gamma_{\alpha\beta}^{m}\Pi_{m}.
\end{align}
When acting on superfields $\Phi=\Phi(X,\theta)$, they lead to
\begin{align}
\Pi^{m}(z)\,\Phi(y) & \sim-\left(\frac{\alpha'}{2}\right)\frac{1}{(z-y)}\partial_{m}\Phi,\\
d_{\alpha}(z)\,\Phi(y) & \sim + \frac{1}{(z-y)}D_{\alpha}\Phi,
\end{align}
with $D_{\alpha}=\frac{\partial}{\partial \theta^{\alpha}}+\zeta(\gamma^{m}\theta)_{\alpha}\frac{\partial}{\partial  X^{m}}$.

The ghost variables are denoted by $\lambda^{\alpha}$, and satisfy
the pure spinor constraint $(\lambda\gamma^{m}\lambda)=0.$ Their
conjugates, $w_{\alpha}$, usually appear through the currents
\begin{align*}
N^{mn} & =-\frac{1}{2}w_{\alpha}\lambda^{\beta}(\gamma^{mn})_{\phantom{\alpha}\beta}^{\alpha},\\
J & =-w_{\alpha}\lambda^{\alpha},
\end{align*}
with $(\gamma^{mn})_{\phantom{\alpha}\beta}^{\alpha}=\frac{1}{2}(\gamma^{m\alpha\rho}\gamma_{\rho\beta}^{n}-\gamma^{n\alpha\rho}\gamma_{\rho\beta}^{m})$,
respectively the Lorentz current in the ghost sector, and the ghost
number current. The relevant OPEs are
\begin{align}
N^{mn}(z)\,\lambda^{\alpha}(y) & \sim + \frac{1}{2}\frac{1}{(z-y)}(\gamma^{mn}\lambda)^{\alpha},\\
J(z)\,\lambda^{\alpha}(y) & \sim + \frac{1}{(z-y)}\lambda^{\alpha},\\
N^{mn}(z)\,N^{pq}(y) & \sim-\frac{3}{(z-y)^{2}}(\eta^{mq}\eta^{np}-\eta^{mp}\eta^{nq})\nonumber \\
 & \phantom{\sim} +\frac{1}{(z-y)}(\eta^{nq}N^{mp}+\eta^{mp}N^{nq}-\eta^{np}N^{mq}-\eta^{mq}N^{np})\label{eq:LLghost-OPE}\\
J(z)\,J(y) & \sim-\frac{4}{(z-y)^{2}}.
\end{align}

Finally, we introduce the (holomorphic) pure spinor BRST charge,
\begin{equation}
Q=\oint\lambda^{\alpha}d_{\alpha},
\end{equation}
which is nilpotent as a consequence of the pure spinor constraint. It is possible
to show that the total energy-momentum tensor, $T$, defined as
\begin{equation}
T=-\frac{1}{\alpha'}\partial X^{m}\partial X_{m}-p_{\alpha}\partial\theta^{\alpha}-w_{\alpha}\partial\lambda^{\alpha},
\end{equation}
is BRST-exact \cite{Berkovits:2004px,Berkovits:2005bt}, and has vanishing central charge.

\subsection{Physical spectrum}

The ghost number one cohomology of $Q$ contains the physical states
of the formalism. At the massless level, the unintegrated vertex operator
is simply
\begin{equation}
U_{\textrm{PS}}=\lambda^{\alpha}A_{\alpha}.\label{eq:un-vertex}
\end{equation}
BRST-closedness leads to
\begin{equation}
\lambda^{\alpha}\lambda^{\beta}D_{\alpha}A_{\beta}=0,
\end{equation}
which is equivalent to the linearized equation of motion of super Yang-Mills in
ten dimensions, encoded in the superfield $A_{\alpha}=A_{\alpha}(X,\theta)$.
Gauge invariance follows the usual BRST-exact identification, which
here translates to $\delta A_{\alpha}=D_{\alpha}\Lambda$, with superfield
parameter $\Lambda$.

The cohomology also contains  the integrated version of the vertex operator \eqref{eq:un-vertex},
whose integrand $V_{\textrm{PS}}$ satisfies $[Q,V_{\textrm{PS}}]=\partial U_{\textrm{PS}}$.
It can be cast as
\begin{equation}
V_{\textrm{PS}}=\partial\theta^{\alpha}A_{\alpha}+\Pi^{m}A_{m}+\frac{\alpha'}{4}d_{\alpha}W^{\alpha}+\frac{\alpha'}{4}N^{mn}F_{mn}.\label{eq:in-vertex}
\end{equation}
The additional superfields in \eqref{eq:in-vertex} are defined by
\begin{align}
\gamma_{\alpha\beta}^{m}A_{m} & \equiv\frac{1}{2\zeta}(D_{\alpha}A_{\beta}+D_{\beta}A_{\alpha}),\\
(\gamma_{m}W)_{\alpha} & \equiv\frac{1}{\zeta}(D_{\alpha}A_{m}-\partial_{m}A_{\alpha}),\\
F_{mn} & \equiv\partial_{m}A_{n}-\partial_{n}A_{m},
\end{align}
and satisfy
\begin{align}
D_{\alpha}W^{\beta} & =-\frac{1}{2}(\gamma^{mn})_{\phantom{\alpha}\alpha}^{\beta}F_{mn},\\
D_{\alpha}F_{mn} & =\partial_{m}(\gamma_{n}W)_{\alpha}-\partial_{n}(\gamma_{m}W)_{\alpha},\\
\gamma_{\alpha\beta}^{m}\partial_{m}W^{\beta} & =0,\\
\partial^{n}F_{mn} & =0.
\end{align}
Notice that $V_\textrm{PS}$ is a conformal primary field as long as we are in
the transverse gauge $\partial^{m}A_{m}=0$. This is akin to the
massless vertex operator in bosonic string theory.

\subsection{Comparison with the spinning string}

The pure spinor superstring has manifest spacetime supersymmetry, unlike the
spinning string. RNS vertex operators of the so-called Ramond sector
involve half-integer pictures and spin fields, introducing technical
complications even when computing  tree level correlators. On the other
hand, vertex operators from the NS sector take a simple form.
For example, the analogue of $V_{\textrm{PS}}$ in \eqref{eq:in-vertex}
for the massless bosons is given by
\begin{equation}
V_{\textrm{RNS}}=\partial X^{m}a_{m}+\frac{\alpha'}{4}L_{\textrm{RNS}}^{mn}f_{mn},\label{eq:in-vertex-RNS}
\end{equation}
where $f_{mn}=\partial_{m}a_{n}-\partial_{n}a_{m}$, and $L_{\textrm{RNS}}^{mn}\equiv-\psi^{m}\psi^{n}$,
with $\psi^{m}$ denoting the world-sheet superpartners of $X^{m}$.
Using the RNS BRST charge,
\begin{equation}
Q_{\textrm{RNS}}=\oint\left\{ c\left(-\frac{1}{\alpha'}\partial X^{m}\partial X_{m}-\frac{1}{2}\psi^{m}\partial\psi_{m}\right)+\gamma\psi_{m}\partial X^{m}+\textrm{ghosts}\right\} ,
\end{equation}
it is straightforward to show that
\begin{align}
[Q_{\textrm{RNS}},V_{\textrm{RNS}}] & =\partial U_{\textrm{RNS}},\\
U_{\textrm{RNS}} & =cV_{\textrm{RNS}}-\frac{\alpha'}{2}\gamma\psi^{m}a_{m}-\frac{\alpha'}{4} \partial c \partial^{m} a_{m}.
\end{align}
with $\partial^m f_{mn}=0$. $U_{\textrm{RNS}}$ is simply
the unintegrated vertex operator of the gluon, with $c$ and $\gamma$ denoting the superconformal ghosts.

As it turns out, the superfield expansion of \eqref{eq:in-vertex}
is similar to \eqref{eq:in-vertex-RNS} when we focus on the bosonic
polarization. For a given gauge choice, the superfield equations of motion lead to
\begin{align}
\left.A_{\alpha}\right|_{\mathrm{boson}} & =\zeta(\gamma^{m}\theta)_{\alpha}a_{m}-\frac{\zeta^{2}}{8}(\gamma_{p}\theta)_{\alpha}(\theta\gamma^{mnp}\theta)f_{mn}+\mathcal{O}(\theta^{5}),\\
\left.A_{m}\right|_{\mathrm{boson}} & =a^{m}-\frac{\zeta}{4}(\theta\gamma^{mnp}\theta)f_{np}+\mathcal{O}(\theta^{4}),\\
\left.W^{\alpha}\right|_{\mathrm{boson}} & =-\frac{1}{2}(\gamma^{mn}\theta)^{\alpha}f_{mn}+\mathcal{O}(\theta^{3}),\\
\left.F_{mn}\right|_{\mathrm{boson}} & =f_{mn}+\mathcal{O}(\theta^{2}),
\end{align}
which imply
\begin{align}
\left.V_{\textrm{PS}}\right|_{\mathrm{boson}} & =\partial X^{m}a_{m}+\frac{\alpha'}{4}L_{\textrm{PS}}^{mn}f_{mn}+\mathcal{O}(\theta^{2}),\label{eq:in-vertex-expansion}\\
L_{\textrm{PS}}^{mn} & \equiv N^{mn}-\frac{1}{2}(p\gamma^{mn}\theta).
\end{align}

The similarity between \eqref{eq:in-vertex-RNS} and \eqref{eq:in-vertex-expansion}
is undeniable. Furthermore, note that $L_{\textrm{RNS}}^{mn}$ and
$L_{\textrm{PS}}^{mn}$ satisfy the same OPE,
\begin{multline}
L^{mn}(z)\,L^{pq}(y)\sim \frac{1}{(z-y)^{2}}(\eta^{mq}\eta^{np}-\eta^{mp}\eta^{nq})\\
+\frac{1}{(z-y)}(\eta^{nq}L^{mp}+\eta^{mp}L^{nq}-\eta^{np}L^{mq}-\eta^{mq}L^{np}).
\end{multline}
The double pole contribution, for instance, has been a hallmark hint
of the equivalence between the two formalisms. The reason is that
the anomaly of the ghost Lorentz algebra in \eqref{eq:LLghost-OPE}
supports the naive identification $L_{\textrm{RNS}}^{mn} \approx L_{\textrm{PS}}^{mn}$.

\section{Tree level correlators}

In this section we review three-point amplitudes in both formalisms,
RNS  and pure-spinor. In RNS, we compute an amplitude with bosons only, while in the pure-spinor case we compute the full SYM amplitude, containing bosons and fermions.

\subsection{Spinning string}

In the RNS formalism, tree-level correlators can be non-zero only
when they saturate the background charges associated to the ghost
variables. From the path integral point-of-view, they are essentially
mapped to the three zero modes of the (fermionic) reparametrization
ghost $c$ and to the two zero modes of the (bosonic) superghost $\gamma$ \cite{Friedan:1985ge,Polchinski:1998rr}.
We will simply take
\begin{equation}
\left\langle c(z_{1})c(z_{2})c(z_{3})\delta(\gamma)\delta(\gamma)\right\rangle =z_{12}z_{23}z_{31}.\label{eq:ghost-measure-RNS}
\end{equation}

Therefore, if we are interested in computing the three-gluon amplitude,
we have to evaluate
\begin{align}
\mathcal{A}_{3}\left(a^{(1)},a^{(2)},a^{(3)}\right) & =\left\langle \delta(\gamma)\delta(\gamma)U_{\textrm{RNS}}^{(1)}U_{\textrm{RNS}}^{(2)}U_{\textrm{RNS}}^{(3)}\right\rangle ,\nonumber \\
 & =z_{12}z_{23}z_{31}\left\langle V_{\textrm{RNS}}^{(1)}V_{\textrm{RNS}}^{(2)}V_{\textrm{RNS}}^{(3)}\right\rangle _{\textrm{matter}}.\label{eq:3pt-RNS-VVV}
\end{align}
From the first to the second line we have just used the ghost measure
in \eqref{eq:ghost-measure-RNS}. The $\textrm{SL}(2,\mathbb{C})$
invariance of the amplitude requires that
\begin{equation}
\left\langle V_{\textrm{RNS}}^{(1)}V_{\textrm{RNS}}^{(2)}V_{\textrm{RNS}}^{(3)}\right\rangle _{\textrm{matter}}\propto\frac{1}{z_{12}z_{23}z_{31}}.
\end{equation}

The explicit computation of the correlator is straightforward. We
start with
\begin{equation}
\left\langle V_{\textrm{RNS}}^{(1)}V_{\textrm{RNS}}^{(2)}V_{\textrm{RNS}}^{(3)}\right\rangle _{\textrm{matter}}=\left\langle \prod_{i=1}^{3}\left(\partial X^{m_i}a_{m_i}^{(i)}+\frac{\alpha'}{4}L_{\textrm{RNS}}^{m_i n_i}f_{m_i n_i}^{(i)}\right)\right\rangle _{X,\psi},
\end{equation}
and evaluate all possible Wick contractions. There are two non-trivial
correlators involving the Lorentz currents:
\begin{equation}
\left\langle L_{\textrm{RNS}}^{mn}(z_{i})L_{\textrm{RNS}}^{pq}(z_{j})\right\rangle _{\psi}= \frac{1}{z_{ij}^{2}}(\eta^{mq}\eta^{np}-\eta^{mp}\eta^{nq}),
\end{equation}
and
\begin{multline}
\left\langle L_{\textrm{RNS}}^{mn}(z_{1})L_{\textrm{RNS}}^{pq}(z_{2})L_{\textrm{RNS}}^{rs}(z_{3})\right\rangle _{\psi}=\\
=\frac{1}{z_{12}z_{23}z_{31}}(\eta^{ms}\eta^{nq}\eta^{pr}+\eta^{mr}\eta^{np}\eta^{qs}+\eta^{mq}\eta^{nr}\eta^{ps}+\eta^{mp}\eta^{ns}\eta^{qr})\\
-\frac{1}{z_{12}z_{23}z_{31}}(\eta^{ms}\eta^{np}\eta^{qr}+\eta^{mr}\eta^{nq}\eta^{ps}+\eta^{mq}\eta^{ns}\eta^{pr}+\eta^{mp}\eta^{nr}\eta^{qs}).
\end{multline}
We then obtain
\begin{align}
\left\langle V_{\textrm{RNS}}^{(1)}V_{\textrm{RNS}}^{(2)}V_{\textrm{RNS}}^{(3)}\right\rangle _{\textrm{matter}} & = \left\langle \partial X^{m}a_{m}^{(1)}(z_{1})\partial X^{n}a_{n}^{(2)}(z_{2})\partial X^{p}a_{p}^{(3)}(z_{3})\right\rangle _{X}\nonumber \\
 & -\left(\frac{\alpha'}{4}\right)^{2}\frac{2}{z_{23}^{2}}\left\langle \partial X^{m}a_{m}^{(1)}(z_{1})f_{np}^{(2)}(z_{2})f^{(3)np}(z_{3})\right\rangle _{X}\nonumber \\
 & -\left(\frac{\alpha'}{4}\right)^{2}\frac{2}{z_{31}^{2}}\left\langle f_{np}^{(1)}(z_{1})\partial X^{m}a_{m}^{(2)}(z_{2})f^{(3)np}(z_{3})\right\rangle _{X}\nonumber \\
 & -\left(\frac{\alpha'}{4}\right)^{2}\frac{2}{z_{12}^{2}}\left\langle f_{np}^{(1)}(z_{1})f^{(2)np}(z_{2})\partial X^{m}a_{m}^{(3)}(z_{3})\right\rangle _{X}\nonumber \\
 & +\left(\frac{\alpha'}{4}\right)^{3}\frac{8}{z_{12}z_{23}z_{31}}\left\langle f_{m}^{(1)n}(z_{1})f_{n}^{(2)p}(z_{2})f_{p}^{(3)m}(z_{3})\right\rangle _{X}.\label{eq:3pt-RNS-VVV expansion}
\end{align}
Now we are left with bosonic string-like correlators, which finally lead to
\begin{multline}
\left\langle V_{\textrm{RNS}}^{(1)}V_{\textrm{RNS}}^{(2)}V_{\textrm{RNS}}^{(3)}\right\rangle _{\textrm{matter}}=\frac{1}{z_{12}z_{23}z_{31}}\\
\times\left(\frac{\alpha'}{2}\right)^{2}\eta^{np}\left(a_{n}^{(1)}\partial^{m}a_{p}^{(2)}a_{m}^{(3)}+a_{m}^{(1)}a_{n}^{(2)}\partial^{m}a_{p}^{(3)}+\partial^{m}a_{p}^{(1)}a_{m}^{(2)}a_{n}^{(3)}\right),
\end{multline}
where momentum conservation is left implicit. Therefore, 
\begin{equation}
\mathcal{A}_{3}\left(a^{(1)},a^{(2)},a^{(3)}\right)=\frac{1}{2}\left(\frac{\alpha'}{2}\right)^{2}f^{(1)mn}a_{m}^{(2)}a_{n}^{(3)}+\textrm{cyclic}(1,2,3),\label{eq:3pt-RNS}
\end{equation}
which matches the three-point super Yang-Mills partial
amplitude with external gluons.

\subsection{Pure spinor superstring (usual prescription)}

In the pure spinor formalism, tree level correlators can be evaluated using
the measure 
\begin{equation}
\Big\langle (\lambda\gamma^{m}\theta)(\lambda\gamma^{p}\theta)(\lambda\gamma^{q}\theta)(\theta\gamma_{mnp}\theta)\Big\rangle =-180\left(\frac{\alpha'}{2\zeta^{2}}\right)^{2}.\label{eq:PS-measure}
\end{equation}
The term inside brackets is the only element in the ghost number three
cohomology. This is similar to the object $c\partial c\partial^{2}c$
from the bosonic string, for instance. The normalization on the right
hand side of this equation can be consistently fixed. We pick that particular one in order
to simply illustrate a direct agreement with the three-point correlator
\eqref{eq:3pt-RNS-VVV}. 

The pure spinor analogue of \eqref{eq:3pt-RNS-VVV} is built out of three insertions of the unintegrated
vertex operator $\eqref{eq:un-vertex}$, which can be cast as
\begin{equation}
\mathcal{A}_{3}=\left\langle \lambda^{\alpha}A_{\alpha}^{(1)}(z_{1})\lambda^{\beta}A_{\beta}^{(2)}(z_{2})\lambda^{\gamma}A_{\gamma}^{(3)}(z_{3})\right\rangle .
\end{equation}
Using the superfield expansion for $A_{\alpha}$, it is easy to
show that we can choose a gauge in which
\begin{equation}
\lambda^{\alpha}A_{\alpha}=\zeta(\lambda\gamma^{m}\theta)a_{m}+\frac{2\zeta^{2}}{3}(\lambda\gamma^{m}\theta)(\theta\gamma_{m}\xi)-\frac{\zeta^{2}}{8}(\lambda\gamma^{p}\theta)(\theta\gamma_{mnp}\theta)f^{mn}+\mathcal{O}(\theta^{4}),
\end{equation}
where $\xi^{\alpha}$ is the spacetime superpartner of $a_{m}$. Given
that, in this gauge, the lowest superfield component is linear in $\theta^{\alpha}$,
the expansion above is enough to saturate the measure \eqref{eq:PS-measure}.
It is straightforward to show that
\begin{multline}
\mathcal{A}_{3}=-\frac{\zeta^{4}}{8}\left\langle (\lambda\gamma^{m}\theta)(\lambda\gamma^{n}\theta)(\lambda\gamma^{r}\theta)(\theta\gamma_{pqr}\theta)\right\rangle f^{(1)pq}a_{m}^{(2)}a_{n}^{(3)}\\
+\frac{4\zeta^{5}}{9}\left\langle (\lambda\gamma^{m}\theta)(\lambda\gamma^{n}\theta)(\gamma_{n}\theta)_{\alpha}(\lambda\gamma^{p}\theta)(\gamma_{p}\theta)_{\beta}\right\rangle a_{m}^{(1)}\xi^{(2)\alpha}\xi^{(3)\beta}+\textrm{cyclic}(1,2,3).\label{eq:3pt-PS-before measure}
\end{multline}

Now we just need the following identities, which can be obtained from SO$(9,1)$ covariance arguments,
\begin{align}
\left\langle (\lambda\gamma^{m}\theta)(\lambda\gamma^{n}\theta)(\lambda\gamma^{r}\theta)(\theta\gamma_{pqr}\theta)\right\rangle  & =\frac{1}{90}(\delta_{p}^{m}\delta_{q}^{n}-\delta_{q}^{m}\delta_{p}^{n})\left\langle (\lambda\gamma^{r}\theta)(\lambda\gamma^{s}\theta)(\lambda\gamma^{t}\theta)(\theta\gamma_{rst}\theta)\right\rangle ,\\
 & =-2\left(\frac{\alpha'}{2\zeta^{2}}\right)^{2}(\delta_{p}^{m}\delta_{q}^{n}-\delta_{q}^{m}\delta_{p}^{n}),\\
\left\langle (\lambda\gamma^{m}\theta)(\lambda\gamma^{n}\theta)(\gamma_{n}\theta)_{\alpha}(\lambda\gamma^{p}\theta)(\gamma_{p}\theta)_{\beta}\right\rangle  & =\frac{1}{160}\gamma^m_{\alpha\beta}\left\langle (\lambda\gamma^{r}\theta)(\lambda\gamma^{n}\theta)(\lambda\gamma^{p}\theta)(\theta\gamma_{rnp}\theta)\right\rangle \\
 & =-\frac{9}{8}\left(\frac{\alpha'}{2\zeta^{2}}\right)^{2}\gamma^m_{\alpha\beta}.
\end{align}
After substituting the above expressions into  equation
\eqref{eq:3pt-PS-before measure}, we finally obtain
\begin{equation}
\mathcal{A}_{3}=\frac{1}{2}\left(\frac{\alpha'}{2}\right)^{2}\left(f^{(1)mn}a_{m}^{(2)}a_{n}^{(3)}-\zeta a_{m}^{(1)}(\xi^{(2)}\gamma^{m}\xi^{(3)})\right)+\textrm{cyclic}(1,2,3),\label{eq:3pt-PS}
\end{equation}
as expected. Indeed, there is plenty of evidence that the pure spinor
amplitudes match the ones computed in the RNS formalism (see \cite{Mafra:2022wml}).

\section{Alternative tree level prescription in the pure spinor formalism}

In \cite{Berkovits:2016xnb}, an alternative method for calculating
pure-spinor superstring scattering amplitudes was introduced. It
was inspired by  the work of Lee and Siegel in \cite{Lee:2006pa},
and involves only ghost number zero vertex operators. In this alternative prescription, the $N$-point tree level amplitude
with massless external states is given by
\begin{equation}
\mathcal{A}_{N}\propto z_{12}z_{23}z_{31}\left\langle V_{\textrm{PS}}(z_{1})V_{\textrm{PS}}(z_{2})V_{\textrm{PS}}(z_{3})\prod_{i=4}^{N}\left(\int dz_{i}\,V_{\textrm{PS}}(z_{i})\right)\right\rangle .\label{eq:N-point-PS-new}
\end{equation}
The idea is to perform all Wick contractions of the non-zero conformal
weight operators and pick the $\theta^{\alpha}=0$ component of the
resulting expression. Given that $\mathcal{A}_{N}$ is written in terms of
superfields, it is supposed to compute any amplitude involving  bosons
or fermions.

Let us look at the simplest case, $N=3$:
\begin{equation}
\mathcal{A}_{3}\propto z_{12}z_{23}z_{31}\Big\langle V_{\textrm{PS}}(z_{1})V_{\textrm{PS}}(z_{2})V_{\textrm{PS}}(z_{3})\Big\rangle .
\end{equation}
This expression resembles equation \eqref{eq:3pt-RNS-VVV} in the
RNS formalism, with the identification of $V_{\textrm{PS}}$ and $V_{\textrm{RNS}}$.
While  gauge invariance is guaranteed there, the consistency of this ad-hoc prescription
for pure spinors is far from obvious. In fact, we shall see that it
fails when external fermions are present.

For the sake of completeness, we will present here the full three-point
calculation. The idea is to compute the correlator
\begin{equation}
I(1,2,3)=\Big\langle V_{\textrm{PS}}(z_{1})V_{\textrm{PS}}(z_{2})V_{\textrm{PS}}(z_{3})\Big\rangle ,\label{eq:VVV-PS}
\end{equation}
using the free field OPEs \eqref{eq:OPEs}. The explicit expression for $V_{\textrm{PS}}$ was given in equation \eqref{eq:in-vertex}, from which it is easy to see
that $I(1,2,3)$ will involve only nine potentially non-vanishing, independent
correlators (plus permutations thereof). They are schematically given by
\begin{eqnarray*}
\left\langle \left(\partial\theta A\right)\left(\Pi A\right)\left(dW\right)\right\rangle , & \left\langle \left(\Pi A\right)\left(\Pi A\right)\left(\Pi A\right)\right\rangle , & \left\langle \left(\Pi A\right)\left(NF\right)\left(NF\right)\right\rangle ,\\
\left\langle \left(\partial\theta A\right)\left(dW\right)\left(dW\right)\right\rangle , & \left\langle \left(\Pi A\right)\left(\Pi A\right)\left(dW\right)\right\rangle , & \left\langle \left(dW\right)\left(NF\right)\left(NF\right)\right\rangle ,\\
\left\langle \left(dW\right)\left(dW\right)\left(dW\right)\right\rangle , & \left\langle \left(\Pi A\right)\left(dW\right)\left(dW\right)\right\rangle , & \left\langle \left(NF\right)\left(NF\right)\left(NF\right)\right\rangle .
\end{eqnarray*}
The remaining correlators,
\begin{eqnarray*}
\left\langle \left(\partial\theta A\right)\left(\partial\theta A\right)\left(\partial\theta A\right)\right\rangle , & \left\langle \left(\partial\theta A\right)\left(\Pi A\right)\left(\Pi A\right)\right\rangle , & \left\langle \left(\Pi A\right)\left(\Pi A\right)\left(NF\right)\right\rangle ,\\
\left\langle \left(\partial\theta A\right)\left(\partial\theta A\right)\left(\Pi A\right)\right\rangle , & \left\langle \left(\partial\theta A\right)\left(dW\right)\left(NF\right)\right\rangle , & \left\langle \left(\Pi A\right)\left(dW\right)\left(NF\right)\right\rangle ,\\
\left\langle \left(\partial\theta A\right)\left(\partial\theta A\right)\left(dW\right)\right\rangle , & \left\langle \left(\partial\theta A\right)\left(NF\right)\left(NF\right)\right\rangle , & \left\langle \left(dW\right)\left(dW\right)\left(NF\right)\right\rangle ,
\end{eqnarray*}
vanish because of residual conformal weight one operators,
namely $\partial\theta^{\alpha}$ or $N^{mn}$, which vanish in the path integral.

\subsection{List of non-trivial contributions}

We will start with the correlators involving the ghost Lorentz current,
which comprise
\begin{equation}
\left\langle N^{mn}(z_{i})N^{pq}(z_{j})\right\rangle =-\frac{3}{z_{ij}^{2}}(\eta^{mq}\eta^{np}-\eta^{mp}\eta^{nq}),
\end{equation}
and
\begin{multline}
\left\langle N^{mn}(z_{1})N^{pq}(z_{2})N^{rs}(z_{3})\right\rangle 
= + \frac{3}{z_{12}z_{23}z_{31}}(\eta^{ms}\eta^{nq}\eta^{pr}+\eta^{mr}\eta^{np}\eta^{qs}+\eta^{mq}\eta^{nr}\eta^{ps}+\eta^{mp}\eta^{ns}\eta^{qr})\\
-\frac{3}{z_{12}z_{23}z_{31}}(\eta^{ms}\eta^{np}\eta^{qr}+\eta^{mr}\eta^{nq}\eta^{ps}+\eta^{mq}\eta^{ns}\eta^{pr}+\eta^{mp}\eta^{nr}\eta^{qs}).
\end{multline}
We can use them to show that\begin{subequations}
\begin{align}
I_{123}^{(1)} & =\left\langle N^{mn}F_{mn}^{(1)}N^{pq}F_{pq}^{(2)}N^{rs}F_{rs}^{(3)}\right\rangle, \nonumber \\
 & =-\frac{24}{z_{12}z_{23}z_{31}}F_{m}^{(1)n}F_{n}^{(2)p}F_{p}^{(3)m},
\end{align}
\begin{align}
I_{123}^{(2)} & =\left\langle \Pi^{r}A_{r}^{(1)}(z_{1})N^{mn}F_{mn}^{(2)}(z_{2})N^{pq}F_{pq}^{(3)}(z_{3})\right\rangle, \nonumber \\
 & = \frac{3\alpha'}{z_{12}z_{23}z_{31}}F_{m}^{(1)n}F_{n}^{(2)p}F_{p}^{(3)m},
\end{align}
in addition to the on-shell vanishing correlator
\begin{equation}
\left\langle d_{\alpha}W^{(1)\alpha}(z_{1})N^{mn}F_{mn}^{(2)}(z_{2})N^{pq}F_{pq}^{(3)}(z_{3})\right\rangle =0.
\end{equation}

Next, we compute
\begin{align}
I_{123}^{(3)} & =\left\langle \Pi^{m}A_{m}^{(1)}(z_{1})\Pi^{n}A_{n}^{(2)}(z_{2})\Pi^{p}A_{p}^{(3)}(z_{3})\right\rangle, \nonumber \\
 & = \frac{1}{2}\left(\frac{\alpha'}{2}\right)^{3}\frac{1}{z_{12}z_{23}z_{31}}F_{m}^{(1)n}F_{n}^{(2)p}F_{p}^{(3)m} \nonumber\\
 & \phantom{=}+\frac{1}{2}\left(\frac{\alpha'}{2}\right)^{2}\frac{1}{z_{12}z_{23}z_{31}}\left(F^{(1)mn}A_{m}^{(2)}A_{n}^{(3)}+\textrm{cyclic}(1,2,3)\right),
\end{align}
which is equivalent to the first term in \eqref{eq:3pt-RNS-VVV expansion},
but with superfields.

There are only two relevant correlators involving $\partial\theta^{\alpha}$,
\begin{align}
I_{123}^{(4)} & =\left\langle \partial\theta^{\alpha}A_{\alpha}^{(1)}(z_{1})\Pi^{m}A_{m}^{(2)}(z_{2})d_{\beta}W^{(3)\beta}(z_{3})\right\rangle, \nonumber \\
 & = \frac{\alpha'}{2}\frac{1}{z_{12}z_{23}z_{31}}A_{\alpha}^{(1)}A_{m}^{(2)}\partial^{m}W^{(3)\alpha},
\end{align}
and
\begin{align}
I_{123}^{(5)} & =\left\langle \partial\theta^{\alpha}A_{\alpha}^{(1)}(z_{1})d_{\beta}W^{(2)\beta}(z_{2})d_{\gamma}W^{(3)\gamma}(z_{3})\right\rangle, \nonumber \\
 & =\frac{1}{z_{13}^{2}z_{23}}\partial_{m}A_{\alpha}^{(1)}(\gamma^{n}\gamma^{m}W^{(2)})^{\alpha}A_{n}^{(3)}-\frac{1}{z_{12}^{2}z_{23}}\partial_{m}A_{\alpha}^{(1)}(\gamma^{n}\gamma^{m}W^{(3)})^{\alpha}A_{n}^{(2)}\nonumber \\
 & \phantom{=}+\frac{1}{z_{13}^{2}z_{12}}D_{\beta}A_{\alpha}^{(1)}W^{(2)\beta}W^{(3)\alpha}-\frac{1}{z_{12}^{2}z_{13}}D_{\beta}A_{\alpha}^{(1)}W^{(2)\alpha}W^{(3)\beta}.\label{eq:I5}
\end{align}

Finally, we have
\begin{align}
I_{123}^{(6)} & =\left\langle \Pi^{m}A_{m}^{(1)}(z_{1})\Pi^{n}A_{n}^{(2)}(z_{2})d_{\alpha}W^{(3)\alpha}(z_{3})\right\rangle ,\nonumber \\
 & = \left(\frac{\alpha'}{2}\right)\left(\frac{1}{z_{12}^{2}z_{13}}D_{\alpha}A_{m}^{(1)}A^{(2)m}W^{(3)\alpha}+\frac{1}{z_{12}^{2}z_{23}}A^{(1)m}D_{\alpha}A_{m}^{(2)}W^{(3)\alpha}\right),\label{eq:I6}
\end{align}
\begin{align}
I_{123}^{(7)} & =\left\langle \Pi^{m}A_{m}^{(1)}(z_{1})d_{\alpha}W^{(2)\alpha}(z_{2})d_{\beta}W^{(3)\beta}(z_{3})\right\rangle ,\nonumber \\
 & =\frac{1}{z_{12}z_{23}z_{31}}\left(2\zeta A_{m}^{(1)}(W^{(2)}\gamma^{m}W^{(3)})-8\left(\frac{\alpha'}{2}\right)F_{m}^{(1)n}F_{n}^{(2)p}F_{p}^{(3)m}\right),\label{eq:I7}
\end{align}
and
\begin{align}
I_{123}^{(8)} & =\left\langle d_{\alpha}W^{(1)\alpha}(z_{1})d_{\beta}W^{(2)\beta}(z_{2})d_{\gamma}W^{(3)\gamma}(z_{3})\right\rangle ,\nonumber \\
 & =\frac{32}{z_{12}z_{23}z_{31}}F_{m}^{(1)n}F_{n}^{(2)p}F_{p}^{(3)m}.
\end{align}
\end{subequations}
Momentum conservation is left implicit in all these
correlators. We have made use of the transverse gauge $\partial^{m}A_{m}^{(i)}=0$
and the superfield equations of motion, since we work with on-shell
states. Note that we have not assumed any zero mode measure
for the superfields.

\subsection{The complete correlator}

The three-vertex correlator \eqref{eq:VVV-PS} can be expressed as
\begin{align}
I(1,2,3) & =\frac{1}{3}\left(\frac{\alpha'}{4}\right)^3I_{123}^{(1)}+\left(\frac{\alpha'}{4}\right)^2I_{123}^{(2)}+\frac{1}{3}I_{123}^{(3)}+\left(\frac{\alpha'}{4}\right)\left(I_{123}^{(4)}+I_{132}^{(4)}\right)+\left(\frac{\alpha'}{4}\right)^2I_{123}^{(5)}\nonumber\\
 & \phantom{=}+\left(\frac{\alpha'}{4}\right)I_{123}^{(6)}+\left(\frac{\alpha'}{4}\right)^2I_{123}^{(7)}+\frac{1}{3}\left(\frac{\alpha'}{4}\right)^3I_{123}^{(8)}+\textrm{cyclic}(1,2,3).
\end{align}
It is then straightforward to gather the different contributions and
to show that
\begin{multline}
I(1,2,3)=\frac{1}{2}\left(\frac{\alpha'}{2}\right)^{2}\frac{1}{z_{12}z_{23}z_{31}}\left(F^{(1)mn}A_{m}^{(2)}A_{n}^{(3)}-\zeta A_{m}^{(1)}(W^{(2)}\gamma^{m}W^{(3)})\right)\\
-\left(\frac{\alpha'}{4}\right)^{2}\left[\partial_{2}\left(\frac{1}{z_{12}z_{31}}\right)G(1,2,3)+\partial_{3}\left(\frac{1}{z_{12}z_{31}}\right)G(1,3,2)\right]+\textrm{cyclic}(1,2,3),\label{eq:I_123}
\end{multline}
where
\begin{equation}
G(1,2,3)\equiv D_{\alpha}A_{\beta}^{(1)}W^{(2)\alpha}W^{(3)\beta}+A_{m}^{(1)}(\gamma^{n}\partial_{n}A^{(2)})^{\alpha}(\gamma^{m}W^{(3)})_{\alpha}.
\end{equation}
In $I(1,2,3)$ we have only the zero modes of $X^{m}$ and $\theta^{\alpha}$
in the superfields. The first line in \eqref{eq:I_123} yields the
expected amplitude when we look at the lowest order in $\theta^{\alpha}$,
but the remaining contributions, containing one boson and two fermions,
break both gauge symmetry and the $\textrm{SL}(2,\mathbb{C})$ invariance
of $\mathcal{A}_{3}$. They were suggestively written as total derivatives
with respect to the fixed coordinates of the vertices, since they
would otherwise be disregarded if the moduli integrations were present.

It is interesting to point out that the terms in the second line of \eqref{eq:I_123} do vanish in the supersymmetric
gauge
\begin{equation}
\gamma^{m\alpha\beta}\partial_{m}A_{\beta}^{(i)}=-\zeta W^{(i)\alpha}+\mathcal{O}(\theta^{2}),
\end{equation}
with
\begin{equation}
D_{\alpha}A_{\beta}^{(i)}=\zeta\gamma_{\alpha\beta}^{m}a_{m}^{(i)}+\mathcal{O}(\theta).
\end{equation}
Indeed, it is easy to show that in this gauge $G(1,2,3)$ vanishes when $\theta^{\alpha}=0$,
therefore decoupling from the amplitude.

This example shows that while the prescription \eqref{eq:N-point-PS-new} is very suggestive
and indeed works for external bosons, it cannot be trusted when spacetime
fermions are involved.

\section{Discussion\label{sec:discussion}}

We have demonstrated in detail that the \emph{alternative} tree level prescription for the pure spinor formalism is not consistent with the decoupling of BRST exact states nor invariant with respect to the position of the fixed vertices. These results were also confirmed via symbolic computation.

Our findings have no implication whatsoever to the usual pure spinor prescription, which has been exhaustively verified at tree level \cite{Mafra:2022wml}. However, they hint at some tension with previous results in the literature, both recent and older, as we discuss in the remainder of this section.

\subsection*{Simpler superstring scattering}

Lee and Siegel's proposal \cite{Lee:2006pa} supported the alternative
prescription \eqref{eq:N-point-PS-new} in the pure spinor formalism.
Given our results, we would like to pinpoint some inconsistencies
in their computation. Naturally, we will focus on the three-point
amplitude involving one boson and two fermions. This computation can
be found in the appendix F of their publication.
In particular, they chose the Wess-Zumino gauge ($A_{\alpha}(X,0)=0$).
It is then easy to identify the correlators contributing to $\mathcal{A}_{3}(a^{(1)},\xi^{(2)},\xi^{(3)})$,
namely \eqref{eq:I5}, \eqref{eq:I6}, and \eqref{eq:I7}, and their
permutations. Notice that the part of the vertex involving the ghost
Lorentz current does not contribute in this case. Below we explicitly
compare our results with equation (F.2) in \cite{Lee:2006pa} multiplied
by $z_{12}z_{23}z_{31}$:
\begin{enumerate}
\item $\left\langle \partial\theta^{\alpha}A_{\alpha}^{(1)}(z_{1})\frac{\alpha'}{4}d_{\beta}W^{(2)\beta}(z_{2})\frac{\alpha'}{4}d_{\gamma}W^{(3)\gamma}(z_{3})\right\rangle $:
\begin{align}
\mathcal{A}_{3}^{'}(a^{(1)},\xi^{(2)},\xi^{(3)})_{\textrm{ours}} & =\left(\frac{\alpha'}{4}\right)^{2}\left(\frac{z_{23}}{z_{31}}+\frac{z_{23}}{z_{12}}\right)\zeta A_{m}^{(1)}(W^{(2)}\gamma^{m}W^{(3)}),\\
\mathcal{A}_{3}^{'}(a^{(1)},\xi^{(2)},\xi^{(3)})_{\textrm{L--S}} & =-\mathrm{i}\left(\frac{z_{23}}{z_{12}}+\frac{z_{23}}{z_{31}}\right)A_{m}^{(1)}(W^{(2)}\gamma^{m}W^{(3)}).
\end{align}
\item $\left\langle \Pi^{m}A_{m}^{(1)}(z_{1})\Pi^{n}A_{n}^{(2)}(z_{2})\frac{\alpha'}{4}d_{\alpha}W^{(3)\alpha}(z_{3})\right\rangle +(2\leftrightarrow3)$:
\begin{align}
\mathcal{A}_{3}^{''}(a^{(1)},\xi^{(2)},\xi^{(3)})_{\textrm{ours}} & =2\left(\frac{\alpha'}{4}\right)^{2}\left(\frac{z_{31}}{z_{12}}+\frac{z_{12}}{z_{31}}\right)\zeta A_{m}^{(1)}(W^{(2)}\gamma^{m}W^{(3)}),\\
\mathcal{A}_{3}^{''}(a^{(1)},\xi^{(2)},\xi^{(3)})_{\textrm{L--S}} & =2\mathrm{i}\left(\frac{z_{31}}{z_{12}}+\frac{z_{12}}{z_{31}}\right)A_{m}^{(1)}(W^{(2)}\gamma^{m}W^{(3)}).
\end{align}
\item $\left\langle \Pi^{m}A_{m}^{(1)}(z_{1})\frac{\alpha'}{4}d_{\alpha}W^{(2)\alpha}(z_{2})\frac{\alpha'}{4}d_{\beta}W^{(3)\beta}(z_{3})\right\rangle $:
\begin{align}
\mathcal{A}_{3}^{'''}(a^{(1)},\xi^{(2)},\xi^{(3)})_{\textrm{ours}} & =2\left(\frac{\alpha'}{4}\right)^{2}\zeta A_{m}^{(1)}(W^{(2)}\gamma^{m}W^{(3)}),\\
\mathcal{A}_{3}^{'''}(a^{(1)},\xi^{(2)},\xi^{(3)})_{\textrm{L--S}} & =\mathrm{i}\left(2\frac{z_{23}}{z_{31}}-\frac{z_{12}}{z_{31}}+2\frac{z_{23}}{z_{12}}-\frac{z_{31}}{z_{12}}\right)A_{m}^{(1)}(W^{(2)}\gamma^{m}W^{(3)}).
\end{align}
\end{enumerate}
If we add the contributions, we obtain the following expressions,
\begin{align}
\mathcal{A}_{3}(a^{(1)},\xi^{(2)},\xi^{(3)})_{\textrm{ours}} & =\left(\frac{\alpha'}{4}\right)^{2}\left(\frac{z_{31}}{z_{12}}+\frac{z_{12}}{z_{31}}\right)\zeta A_{m}^{(1)}(W^{(2)}\gamma^{m}W^{(3)}),\label{eq:3pt-bff}\\
\mathcal{A}_{3}(a^{(1)},\xi^{(2)},\xi^{(3)})_{\textrm{L--S}} & =-2\mathrm{i}A_{m}^{(1)}(W^{(2)}\gamma^{m}W^{(3)}),
\end{align}
which clearly disagree, irrespective of the overall normalization.
This mismatch is traced back to two correlators. We can identify (1)
an overall factor in $\mathcal{A}_{3}^{'}$ , and (2) a different
$z_{ij}$ dependence in $\mathcal{A}_{3}^{'''}$. The latter follows
from the free-field correlator $\left\langle \Pi^{m}(z_{1})d_{\alpha}(z_{2})d_{\beta}(z_{3})\right\rangle $.
The correct result is
\begin{equation}
\left\langle \Pi^{m}(z_{1})d_{\alpha}(z_{2})d_{\beta}(z_{3})\right\rangle =-2\zeta\gamma_{\alpha\beta}^{m}\frac{1}{z_{12}z_{23}z_{31}},
\end{equation}
which disagrees with the $z_{ij}$ dependence of the corresponding
correlator in equation (2.1) of \cite{Lee:2006pa}.

\subsection*{B-RNS-GSS formalism}

Recently, a new description of the superstring was introduced, suggesting a unified framework for the spinning string, the Green-Schwarz superstring, and the pure spinor formalism  \cite{Berkovits:2021xwh}. It has been dubbed the B-RNS-GSS formalism.

The world-sheet of the B-RNS-GSS superstring can be seen as a non-minimal version of the spinning string that includes a pure spinor. It contains the usual reparametrization ghosts plus world-sheet superpartners. In addition, it contains   the spacetime superpartners of the target space coordinates. In particular, the massless vertex operators, unintegrated and integrated, can be written as
\begin{align}
    U & = c V_\textrm{PS} - \gamma \psi^m A_m + \ldots, \label{eq:XYZ-vertex}\\
    V & \equiv b_{-1}\cdot U, \nonumber \\
     & = V_\textrm{PS},
\end{align}
respectively, where the ellipsis in $U$ contains the remaining terms needed for BRST-closedness, involving RNS-like spin fields and bosonized ghosts. Notice, however, that those extra terms appear to decouple from $N$-point tree level amplitudes, which would  be naively defined as
\begin{equation} \label{eq:N-pointXYZ}
\mathcal{A}_{N} = \left\langle U(z_{1}) U(z_{2}) U(z_{3})\prod_{i=4}^{N}\left(\int dz_{i}\,V (z_{i})\right)\right\rangle .
\end{equation}
We are leaving implicit the integration measure for the ghosts. Given that we need to saturate the three zero modes from the $c$ ghost, this formula would lead back to \eqref{eq:N-point-PS-new}.

Therefore, our results demonstrate that the tree level amplitude in the B-RNS-GSS formalism, as initially proposed, is incomplete. At the present stage, we can only speculate on the solution, but it is probably related to a non-trivial interplay between the ghost fields in the vertex \eqref{eq:XYZ-vertex} and the saturation of the background ghost charges in the conformal field theory. For example, a naive correction to the vertex $U$ in \eqref{eq:XYZ-vertex} (equation (2.16) in \cite{Berkovits:2021xwh}) of the form
\begin{equation}
    \delta U =  \frac{c}{\gamma} \psi^m (\partial c B_m +  \Lambda^\alpha B_{m \alpha}),
\end{equation}
where $\{B_m,B_{m \alpha}\}$ are superfields\footnote{The inverse power of $\gamma$ looks problematic, but it disappears when saturating the ghost measure in the tree level correlator.}, would lead to correlator contributions in \eqref{eq:N-pointXYZ} beyond \eqref{eq:N-point-PS-new}. The solution to this problem will be left for future investigations, but we are optimistic it can be found, in particular because of the richer world-sheet content available.

\paragraph*{Acknowledgments}
We would like to thank Nathan Berkovits, Humberto Gomez, and Carlos Mafra for useful discussions. RLJ is supported by the GA\v{C}R grant 25-16244S from the Czech Science Foundation. The work of TA was partially supported by the European Structural and Investment Funds and the Czech Ministry of Education, Youth and Sports (project
{\tt FORTE CZ.02.01.01/00/22\_008/0004632}), through its research mobility program. AG is supported by a Royal Society funding, URF\textbackslash{R}\textbackslash221015. SPK is supported by the Marie Sk\l odowska-Curie Actions -- COFUND project, which is co-funded by the European Union (Physics for Future -- Grant Agreement No. 101081515). SPK would like to thank ICTP-SAIFR (FAPESP grant 2021/14335-0) where part of this work was done.

\appendix  

\section{Simple check}

In this appendix we present a simple check of the three-point computation in the alternative prescription, with one boson and two fermions.
The vertex for the boson is given in equation \eqref{eq:in-vertex-expansion}. A similar expression can be found for the fermion. In the Wess-Zumino gauge, it is given by
\begin{equation}
\left.V_{\textrm{PS}}\right|_{\mathrm{fermion}} = \frac{\alpha'}{4} p_\alpha \xi^\alpha + \frac{\zeta}{2} \partial X^m (\theta \gamma_m \xi) + \mathcal{O}(\theta^{2}).
\end{equation}
Higher order powers in $\theta$ are unimportant in our analysis.

We want to compute the three-point correlator
\begin{equation}
\left\langle bff\right\rangle = \left\langle V^{(1)}_{\textrm{PS}}(z_1) V^{(2)}_{\textrm{PS}}(z_2)V^{(3)}_{\textrm{PS}}(z_3)\right\rangle,
\end{equation}
in which particle 1 has a polarization vector $a^{(1)}_m$, and particles 2 and 3 have polarization spinors $\xi^{(2) \alpha}$ and $\xi^{(3) \alpha}$, respectively. We have
\begin{multline}
\left\langle bff\right\rangle = \bigg<\left(\partial X^{m}a_{m}^{(1)}+\frac{\alpha'}{4}L_{\textrm{PS}}^{mn}f_{mn}^{(1)}\right)(z_{1})\\
\times\left(\frac{\alpha'}{4}p_{\alpha}\xi^{(2)\alpha}+\frac{\zeta}{2}\partial X^{n}(\theta\gamma_{n}\xi^{(2)})\right)(z_{2})\left(\frac{\alpha'}{4}p_{\beta}\xi^{(3)\beta}+\frac{\zeta}{2}\partial X^{p}(\theta\gamma_{p}\xi^{(3)})\right)(z_{3})\bigg>+\mathcal{O}(\theta).
\end{multline}
The term involving the Lorentz generator does not contribute at the order $\theta=0$. Each $p_\alpha$ from the fermion vertices has to be contracted with the term linear in $\theta$ from the other fermion vertex. Therefore, the relevant contributions are
\begin{multline}
\left\langle bff\right\rangle = \frac{\zeta\alpha'}{8}\bigg<\left(\partial X^{m}a_{m}^{(1)}\right)(z_{1})\left(p_{\alpha}\xi^{(2)\alpha}\right)(z_{2})\left(\partial X^{p}(\theta\gamma_{p}\xi^{(3)})\right)(z_{3})\bigg>\\
+\frac{\zeta\alpha'}{8}\bigg<\left(\partial X^{m}a_{m}^{(1)}\right)(z_{1})\left(\partial X^{n}(\theta\gamma_{n}\xi^{(2)})\right)(z_{2})\left(p_{\beta}\xi^{(3)\beta}\right)(z_{3})\bigg>+\mathcal{O}(\theta),
\end{multline}
which leads to \eqref{eq:3pt-bff} at the lowest order in the superfields.


\begin{thebibliography}{99}

\bibitem{Ramond:1971gb}P.~Ramond,
``Dual Theory for Free Fermions,''
Phys.\ Rev.\ D {\bf 3}, 2415 (1971).
doi:10.1103/PhysRevD.3.2415

\bibitem{Neveu:1971rx}A.~Neveu and J.~H.~Schwarz,
``Factorizable dual model of pions,''
Nucl.\ Phys.\ B {\bf 31}, 86 (1971).
doi:10.1016/0550-3213(71)90448-2

\bibitem{Green:1980zg}M.~B.~Green and J.~H.~Schwarz,
``Supersymmetrical Dual String Theory,''
Nucl.\ Phys.\ B {\bf 181}, 502 (1981).
doi:10.1016/0550-3213(81)90538-1

\bibitem{Green:1981yb}M.~B.~Green and J.~H.~Schwarz,
``Supersymmetrical String Theories,''
Phys.\ Lett.\  {\bf 109B}, 444 (1982).
doi:10.1016/0370-2693(82)91110-8

\bibitem{Berkovits:2000fe}N.~Berkovits,
``Super Poincare covariant quantization of the superstring,''
JHEP {\bf 0004}, 018 (2000)
doi:10.1088/1126-6708/2000/04/018
[hep-th/0001035].

\bibitem{Mafra:2018nla}
C.~R.~Mafra and O.~Schlotterer,
``Towards the n-point one-loop superstring amplitude. Part I. Pure spinors and superfield kinematics,''
JHEP \textbf{08} (2019), 090
doi:10.1007/JHEP08(2019)090
[arXiv:1812.10969 [hep-th]].

\bibitem{Mafra:2018pll}
C.~R.~Mafra and O.~Schlotterer,
``Towards the n-point one-loop superstring amplitude. Part II. Worldsheet functions and their duality to kinematics,''
JHEP \textbf{08} (2019), 091
doi:10.1007/JHEP08(2019)091
[arXiv:1812.10970 [hep-th]].

\bibitem{Mafra:2018qqe}
C.~R.~Mafra and O.~Schlotterer,
``Towards the n-point one-loop superstring amplitude. Part III. One-loop correlators and their double-copy structure,''
JHEP \textbf{08} (2019), 092
doi:10.1007/JHEP08(2019)092
[arXiv:1812.10971 [hep-th]].

\bibitem{Berkovits:2005ng}N.~Berkovits and C.~R.~Mafra,
``Equivalence of two-loop superstring amplitudes in the pure spinor and RNS formalisms,''
Phys.\ Rev.\ Lett.\  {\bf 96}, 011602 (2006)
doi:10.1103/PhysRevLett.96.011602
[hep-th/0509234].

\bibitem{Gomez:2013sla}
H.~Gomez and C.~R.~Mafra,
``The closed-string 3-loop amplitude and S-duality,''
JHEP \textbf{10} (2013), 217
doi:10.1007/JHEP10(2013)217
[arXiv:1308.6567 [hep-th]].

\bibitem{Berkovits:2002qx}
N.~Berkovits and O.~Chandia,
``Massive superstring vertex operator in D = 10 superspace,''
JHEP \textbf{08} (2002), 040
doi:10.1088/1126-6708/2002/08/040
[arXiv:hep-th/0204121 [hep-th]].

\bibitem{Chakrabarti:2017vld}
S.~Chakrabarti, S.~P.~Kashyap and M.~Verma,
``Theta Expansion of First Massive Vertex Operator in Pure Spinor,''
JHEP \textbf{01} (2018), 019
doi:10.1007/JHEP01(2018)019
[arXiv:1706.01196 [hep-th]].

\bibitem{Chakrabarti:2018mqd}
S.~Chakrabarti, S.~P.~Kashyap and M.~Verma,
``Integrated Massive Vertex Operator in Pure Spinor Formalism,''
JHEP \textbf{10} (2018), 147
doi:10.1007/JHEP10(2018)147
[arXiv:1802.04486 [hep-th]].

\bibitem{Cangemi:2022abk}
L.~Cangemi and P.~Pichini,
``Classical limit of higher-spin string amplitudes,''
JHEP \textbf{06} (2023), 167
doi:10.1007/JHEP06(2023)167
[arXiv:2207.03947 [hep-th]].

\bibitem{Azevedo:2024rrf}
T.~Azevedo, D.~E.~A.~Matamoros and G.~Menezes,
``Compton scattering from superstrings,''
JHEP \textbf{01} (2025), 140
doi:10.1007/JHEP01(2025)140
[arXiv:2403.08899 [hep-th]].

\bibitem{Firrotta:2024fvi}
M.~Firrotta, E.~Kiritsis and V.~Niarchos,
``Scattering, absorption and emission of highly excited strings,''
JHEP \textbf{01} (2025), 051
doi:10.1007/JHEP01(2025)051
[arXiv:2407.16476 [hep-th]].

\bibitem{Alessio:2025nzd}
F.~Alessio, P.~Di Vecchia, M.~Firrotta and P.~Pichini,
``Searching for Kerr in string amplitudes,''
[arXiv:2506.15529 [hep-th]].

\bibitem{Schlotterer:2010kk}
O.~Schlotterer,
``Higher Spin Scattering in Superstring Theory,''
Nucl. Phys. B \textbf{849} (2011), 433-460
doi:10.1016/j.nuclphysb.2011.03.026
[arXiv:1011.1235 [hep-th]].

\bibitem{Chakrabarti:2018bah}
S.~Chakrabarti, S.~P.~Kashyap and M.~Verma,
``Amplitudes Involving Massive States Using Pure Spinor Formalism,''
JHEP \textbf{12} (2018), 071
doi:10.1007/JHEP12(2018)071
[arXiv:1808.08735 [hep-th]].




\bibitem{Kashyap:2024qor}
S.~P.~Kashyap, C.~R.~Mafra, M.~Verma and L.~Ypanaqu{\'e},
``Massless representation of massive superfields and tree amplitudes with the pure spinor formalism,''
JHEP \textbf{02} (2025), 215
doi:10.1007/JHEP02(2025)215
[arXiv:2407.02436 [hep-th]].

\bibitem{Mafra:2024fiy}
C.~R.~Mafra,
``Towards massive field-theory amplitudes from the cohomology of pure spinor superspace,''
JHEP \textbf{11} (2024), 045
doi:10.1007/JHEP11(2024)045
[arXiv:2407.11849 [hep-th]].

\bibitem{Mafra:2025pmz}
C.~R.~Mafra,
``The massive one-loop four-point string amplitude in pure spinor superspace,''
[arXiv:2508.09930 [hep-th]].

\bibitem{Mafra:2022wml}C.~R.~Mafra and O.~Schlotterer,
``Tree-level amplitudes from the pure spinor superstring,''
Phys. Rept. \textbf{1020} (2023), 1-162
doi:10.1016/j.physrep.2023.04.001
[arXiv:2210.14241 [hep-th]].

\bibitem{Mafra:2025abc}C.~Huang, C.~R.~Mafra, and Y.~X.~Tao,
to appear.

\bibitem{Berkovits:2016xnb}N.~Berkovits,
``Untwisting the pure spinor formalism to the RNS and twistor string in a flat and AdS$_{5} \times$ S$^{5}$ background,''
JHEP \textbf{06} (2016), 127
doi:10.1007/JHEP06(2016)127
[arXiv:1604.04617 [hep-th]].

\bibitem{Lee:2006pa}K.~Lee and W.~Siegel,
``Simpler superstring scattering,''
JHEP \textbf{06} (2006), 046
doi:10.1088/1126-6708/2006/06/046
[arXiv:hep-th/0603218 [hep-th]].

\bibitem{Berkovits:2021xwh}
N.~Berkovits,
``Manifest spacetime supersymmetry and the superstring,''
JHEP \textbf{10} (2021), 162
doi:10.1007/JHEP10(2021)162
[arXiv:2106.04448 [hep-th]].

\bibitem{Berkovits:2004px}
N.~Berkovits,
``Multiloop amplitudes and vanishing theorems using the pure spinor formalism for the superstring,''
JHEP \textbf{09} (2004), 047
doi:10.1088/1126-6708/2004/09/047
[arXiv:hep-th/0406055 [hep-th]].

\bibitem{Berkovits:2005bt}
N.~Berkovits,
``Pure spinor formalism as an N=2 topological string,''
JHEP \textbf{10} (2005), 089
doi:10.1088/1126-6708/2005/10/089
[arXiv:hep-th/0509120 [hep-th]].

\bibitem{Friedan:1985ge}
D.~Friedan, E.~J.~Martinec and S.~H.~Shenker,
``Conformal invariance, supersymmetry and string theory,''
Nucl. Phys. B \textbf{271} (1986), 93-165
doi:10.1016/S0550-3213(86)80006-2

\bibitem{Polchinski:1998rr}
J.~Polchinski,
``String theory. Vol. 2: Superstring theory and beyond,''
Cambridge University Press, 2007,
ISBN 978-0-511-25228-0, 978-0-521-63304-8, 978-0-521-67228-3
doi:10.1017/CBO9780511618123

\end{thebibliography}
\end{document}